# MANAGING INFORMATION AND KNOWLEDGE SHARING CULTURES IN HIGHER EDUCATION INSTITUTIONS

**Leon Andretti Abdillah**

Information Systems Study Program
Computer Science Faculty
Bina Dama University
leon.abdillah@yahoo.com

*Abstracts: Information and knowledge (IK) are very important for any institution including education higher institution. Those IK are stored in every single individual in organization in the form of experiences, skills, etc. The growth of the higher education institution nowadays relies on how an institution manage the dissemination of those IK over the organization by using information technology (IT). This article discusses several ways and tools for engaging persons in the organization to build sharing cultures. This article gives some view of freely availabe application over internet to be used for IK sharing cultures.*

*Keywords: Information and Knowledge Sharing Cultures, Email, eLearning, Blog.*

## 1. Introduction

Information Technology and Communication (ICT) become the main backbone for many aspects in our daily live nowadays. ICT has influenced and changed the dissemination of information and knowledge around the world. Information and knowledge move to a new importance level and become important resources [1]. Wireless and virtual communications between people and organizations nowadays are common and general style. Individuals must work together to collect, analyze, synthesize and disseminate information throughout the work process [2]. It means every person in an organization should interact each others to develop mutual benefit for institution growth.

The implementation of knowledge management (KM) has attracted many researchers concerns. The most popular of knowledge conversion or process is introduced by Nonaka. Nonaka identified four major knowledge conversion into four main dimensions, known as "SECI" [3]: 1) Socialization, 2) Externalization, 3) Combination, and 4) Internalization (figure 1).

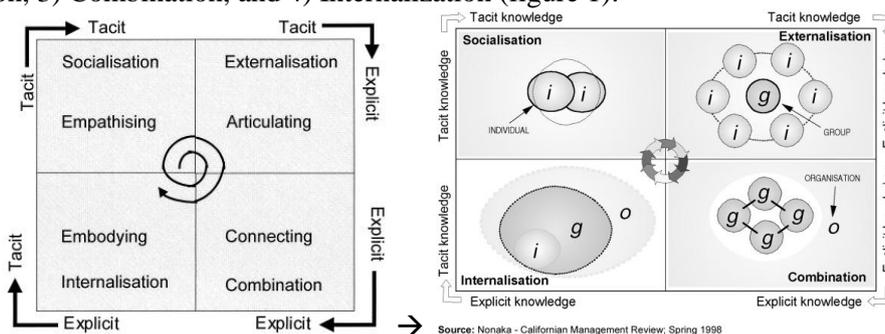

**Figure 1.** SECI process

Author interested to describe the implementation of IT in higher education institution to support information and knowledge sharing. Knowledge are viewed as the principle source of value creation and sustainable competitive advantage [4]. And ICT infrastructure is a key factor of knowledge management effectiveness in project environments and has in some cases been underestimated in previous research [5]. The objective of knowledge management systems (KMS) is to support construction, sharing and application of knowledge in organizations [4]. These organizations could be in education, business, health sector, or banking. But for the domain of this paper, author will focus on education institution.

In this paper, author interest to discuss information and knowledge sharing cultures in higher education institutions. Author defines the term of information and knowledge sharing as the dissemination of information and knowledge from source(s) to destinations(s). The sources could be person or individual or group in an organization. The destinations could also similar to sources but they might be from the same organization or from outside of the particular organization.

Core activities of higher education are associated with knowledge creation and dissemination and learning [6]. In Indonesia, those activities are called "Tri Darma Perguruan Tinggi". It means college or university in Indonesia need to actively participate in three main sectors of education: 1) Education and Teaching (Pendidikan dan Pengajaran), 2) Research and Development (Penelitian dan Pengembangan), 3) Public Services (Pengabdian pada Masyarakat), and followed by 4) Supporting Activities (Unsur Penunjang).

Recent trends in Indonesian higher educations are the use of Information Technology (IT) for every activities. Paperless office become reality in current higher education institutions. Many applications are used for activities: electronic mail, electronic learning, web based documentation, blogging, social networking.

Knowledge sharing opens the oppurtunity to a member of a group, organization, institution or company to share the knowledge, techniques, experiences, and ideas they have to others members. This article discuss how higer education institution starts to engage their community to implement knowledge sharing behavioral as daily cultures. The rest of this paper will discuss methods (section 2), discussions (section 3), and conclusions (section 4).

## 2. Methods

This research uses study literatures and observation methods to gather the information and data. Author involves some students and lecturers to participate in the research. Author also actively involves in *Focus Group Discussion (FGD)* to identify close related aspects about knowledge sharing cultures. This research is conducted during 2013. Then qualitative analysis are described the knowledge sharing tools and cultures. For secondary data collection by using the literature studies and documentation of data organization.

## 3. Discussions

In this section, author suggests several tools combined with classical face-to-face meeting to manage the information and knowledge sharing cultures for higer education institution. Beside weekly meeting, author also include Email, Elearning, Blog, Documentation, Social Network, etc.

### 3.1 Classical Meeting

Even the discussion in this paper are dominantly by IT based activities, but author still consider to include classical weekly meeting as important point of view for semi formal meeting through organization (figure 2).

Every division in institusion needs to disseminate the most recently issues related to their department. Those information basically about Tri Darma Perguruan Tinggi. Sometimes, if there are no urgent condition related to Tri Darm Perguruan Tinggi matter, then every one are able to share their information or knowledge.

Weekly face to face meeting are divided into two groups: 1) Early week will be started by global briefing to discuss various trends and knowledge related to higher education, 2) At thend of the week there is an optional meeting for coordination to discuss urgent conditions or to cover immediate information.

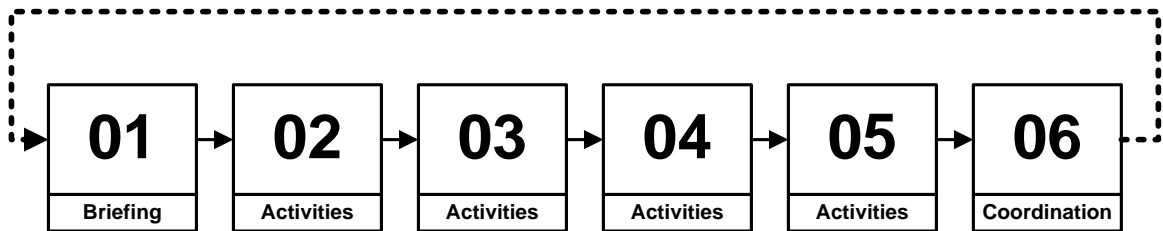

**Figure 2**. Classical Meeting

### 3.2 Documentation

Basically every lecturer has activity related to "Tri Darma Perguruan Tinggi". The evidence of lecturers' activities are very important not only for the lecturers but also for the institution. Those activities need to be stored and organize in close format to DIKTI's standard. The example for this purpose is shown in figure 3.

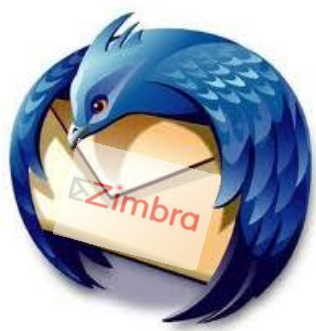

**Figure 3.** Human Resource Information Systems

### 3.3 Electronic Mail (Email)

Email nowadays become an essential media to distribute information among persons in an organization. email is considered as a primary medium for information exchange [7]. In this article author colaborates zimbra and thunderbird.

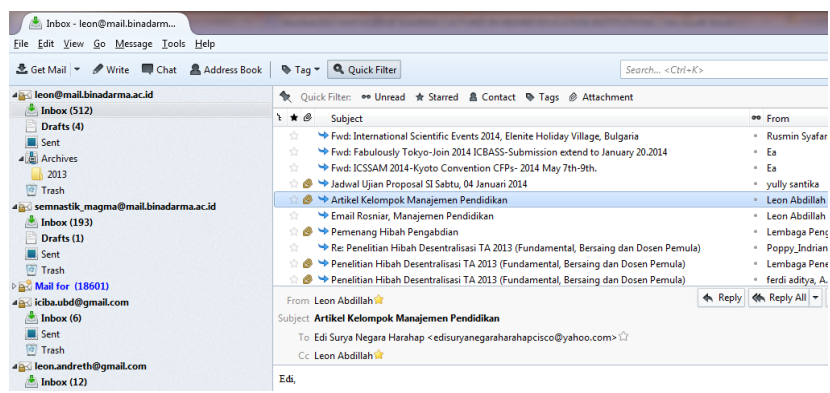

**Figure 4**. A Zimbra email example by suing Thunderbird

Zimbra (www.zimbra.com) is an enterprise-class email, calendar and collaboration solution, built for the cloud, both public and private. With a redesigned browser-based interface, Zimbra offers the most innovative messaging experience available today, connecting end users to the information and activity in their personal clouds.

Thunderbird (www.mozilla.org/en-US/thunderbird) is a free email application that's easy to set up and customize - and it's loaded with great features! The example of zimbra in thunderbird could be seen in figure 4.

In this paper, author discuss the utilization of zimbra to disseminate various information related to "Tri Darma Perguruan Tinggi" plus others administrative activities. Sub divisions are able to share their information. Author it self uses this media to coordinate with conference team to process the submitted papers for future proceeding.

### 3.4 Electronic Learning (eLearning)

Current technology has changed the scheme of clasical learning process and styles. One obvious use of IT to enable knowledge management in through e-learning, the creation and distribution of knowledge through the online delivery of information, communication, education, and training [8]. IT offers many benefits to learning systems. It has changed the way of learning styles and approaches [9]. IT creates paperless evironment by engage the power of PDF [10] as global document format. Major benefits of e-learning included the ease of access to resources, and the provision of central area for students to access to find information or comprehensive resources pertaining to each module [11].

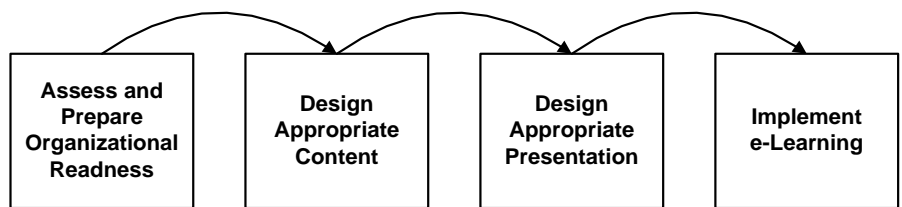

**Figure 5.a.** E-learning value chain [8]

The software used for e-learning is Moodle [12]. Moodle (moodle.org) is a Course Management System (CMS), also known as a Learning Management System (LMS) or Virtual Learning Environment (VLE). It's a free web application that educators can use to create effective online learning sites.

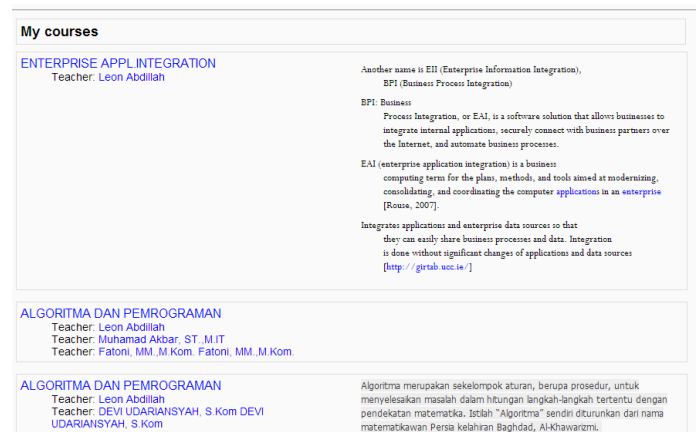

**Figure 5.b.** My course in E-learning main menu

### 3.5 Blog

Weblog (blog) helps as the media for disseminating the knowledge through internet. Blog is design as personal web, where the author of the blog commonly recognize as blogger. Blog become one of the most popular icons in internet application nowadays. Author uses Wordpress (www.wordpress.com)

application to [9] disseminate publications related to "Tri Darma Perguruan Tinggi" such as: 1) knowledge (learning materials) to the students via blog (figure 6), 2) Research publications, 3) public services activities, and 4) supporting activities.

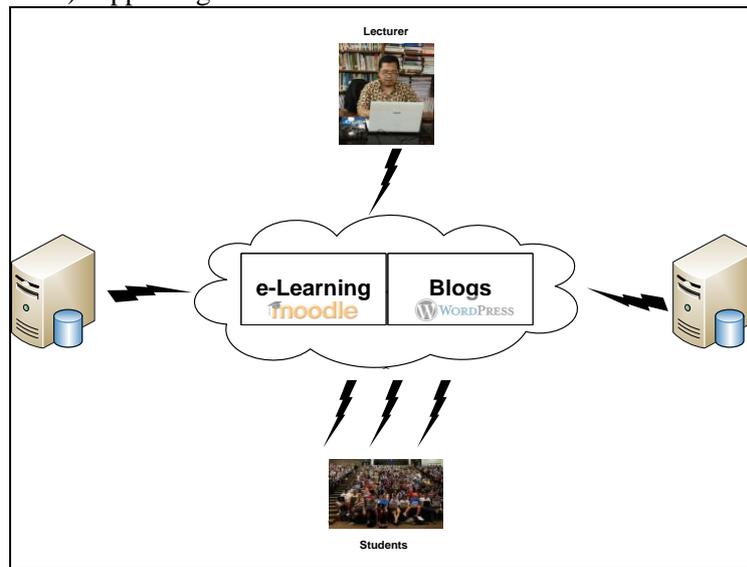

**Figure 6.** Blended e-Learning scheme with blog

### 3.6 Social Networks

Social networking media, is a media that is widely used to access the information [13]. At the moment, social media like facebook and twitter dominantly used by many people around the world. McQuail claimed as new media, social networking has several advantages [14], such as: 1) *Interactivity*, 2) *Social presence (sociability),* 3) *Media richness*, 4) *Autonomy,* 5) *Playfulness,* and 6) *Personalization.* Author adopts the power of social media to engage and disseminate research results, teaching materials, or any academic activities to students and world wide audiens. Author also create several group in social media to get in touch with similar audiens. These Internet tools reduce the cost of social exchanges—social interaction is much easier; and the variety of platforms provides increased flexibility in how we communicate [15]. The example of research and teaching material as post in facebook could be seen in figure 7.

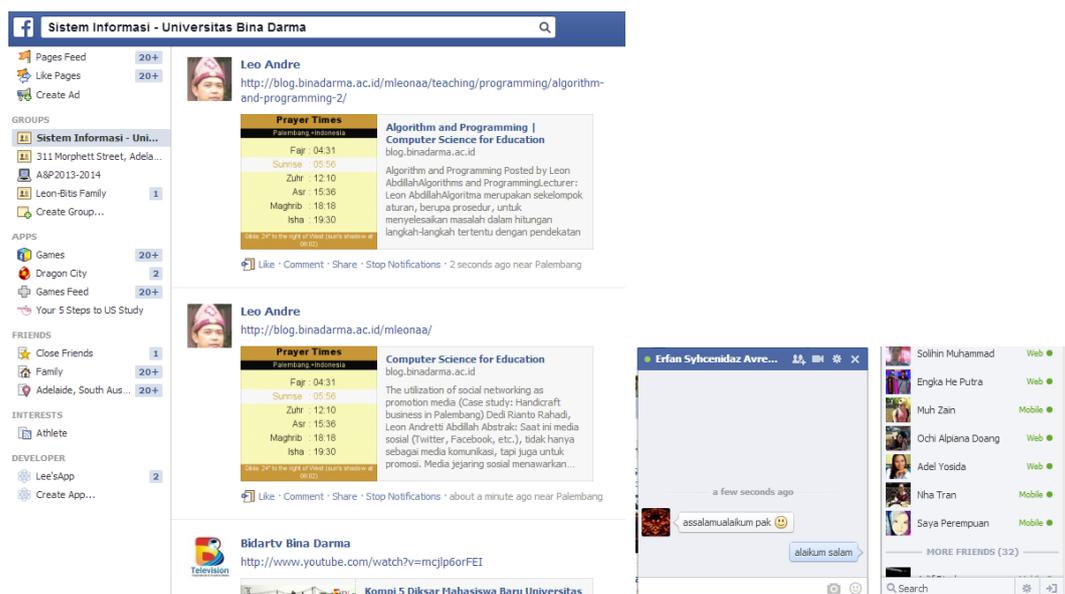

**Figure 7.** Social Network

## 3.7 Repository Software

Academic Institutions also need to store their published documents in repository software. In this article authos discuss the repository software of ePrints (www.eprints.org). Eprints supports research papers, theses, teaching materials, arts and many more. Eprints seems to be a very good system for hosting an incrementally growing collection, such as a faculty publications collection or an electronic theses collection [16]. At the moment, the collections of students and lecturers articles are store in ePrints. The example of eprints could be seen in figure 8.

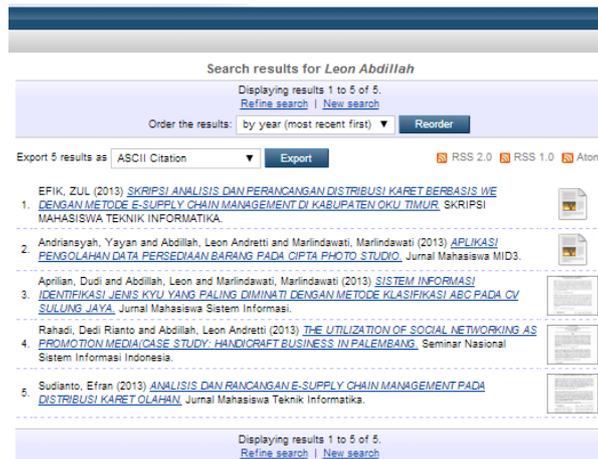

**Figure 8**. An example of Repository Software by using ePrints

## 4. Conclusions

Based on the discussions above, author concludes this article as follows:
   a) For early stage of knowledge management in higer education institution not always involve complicated systems, but institutions could used freely availabe applications over the internet.
   b) As the main key success factor to integrated sparse information over an organiation, IT plays the significant role and become a primary backbone for IT based organizations in the future.
   c) The commitment from top management are vital to support the continuality of knowledge sharing cultures over higher education institutions.
   d) Some higher education institutions need to explore their typical sharing culture as the characteristic and unique value of their orgaization.
   e) For future research, there is a great opportunity for education institution to integrated many applications over the main gate.